# Semiconducting Carbon Nanotubes in Photovoltaic Blends: the case of PTB7:PC$_{60}$BM:(6,5) SWNT

**Diana Gisell Figueroa del Valle**[b§], **Giuseppe M. Paternò**[a§*], **Francesco Scotognella**[b,c]

[a] Instituto Superior de Tecnologias y Ciencias Aplicadas, InSTEC, Ave. Salvador Allende, esq. Luaces, La Habana, Cuba
[b] Center for Nano Science and Technology@PoliMi, Istituto Italiano di Tecnologia, Via Pascoli 70/3, 20133 Milano, Italy
[c] Dipartimento di Fisica, Politecnico di Milano, Piazza L. da Vinci 32, 20133 Milano, Italy

## Abstract

Blends of carbon nanotubes with conjugated polymer and fullerene derivatives are complex nanocomposite systems, which have recently attracted a great research interest for their photovoltaic ability. Therefore, gaining a better understanding of the excitonic dynamics in such materials can be important to boost the efficiency of excitonic solar cells. Here, we studied the photophysics of a ternary system in which the polymer PTB7 and the fullerene derivative PCBM are integrated with (6,5) SWNTs. We highlight the contribution of SWNTs in the exciton dissociation and in the charge transfer process. These findings can be useful for the exploitation of these multi-component systems for organic photovoltaic and, in general, optoelectronic applications.

**Keywords**: carbon nanotubes, excitons, charge-transfer, ultrafast spectroscopy, organic photovoltaic

# 1 Introduction

The concept of excitons in semiconducting carbon nanotubes was firstly introduced by T. Ando in a pioneering theoretical work published in 1997.[1] The idea of optical resonances due to excitons in carbon nanotubes was then reprised in 2004 in an experimental work by the Vardeny and Baughman groups,[2] and a theoretical work by the Louie group.[3] In May 2005 the group of Tony Heinz,[4] and in December 2005 the group of Christoph Lienau,[5] found the smoking gun evidence to state that "the optical resonances in carbon nanotubes arise from excitons", with an exciton binding energy of 420 meV for (7,5), (6,5) and (8,3) single walled carbon nanotubes (SWNTs).[4] Finally, in 2008 the exciton size and mobility for semiconducting carbon nanotubes with (6,5) chirality were measured.[6] Such huge research effort in understanding the fundamental photoexcitation phenomena in carbon nanotubes demonstrates that the exciton picture of carbon nanotubes is in-fact a complex scenario, involving a series of bright and dark excitons as well as higher energetic excitons lying above the bottom of the conduction band.[7,8] Such higher excitons, in particular, can serve as probes for photogenerated charge carriers in the highly enriched (6,5) SWNT (see figure 1a).[9-11]

The employment of SWNTs as active materials in organic photovoltaic diodes (OPVs) is very promising because of their optical absorption in the near infrared region and the relatively high carrier mobility.[12-16] Given the excitonic character of OPVs it is necessary a proper electron donor/acceptor interface to achieve effective excitons splitting and charge generation. Therefore, in organic solar cells SWNTs are usually blended with appropriate electron donor/acceptor systems either in bilayer or bulk heterojunction configurations, i.e. with conjugated polymers and fullerene derivatives. However, the relatively high level of complexity of these blends and the strong dependency of their nanomorphology on the intrinsic structural properties of the constituent materials (i.e. crystallinity)[17-19] and on a number of process parameters (i.e. deposition techniques and casting solvent)[20-22], hinder a complete and reliable characterization of these functional nanocomposites. In this scenario, photophysical studies, as the ones reported for SWNTs in blends with poly(3-hexylthiophene) (P3HT) and phenyl-$C_{61}$-butyric acid methyl ester (PCBM),[23, 24] can give a strong support to describe these systems, with the view to improve their photogeneration ability.

Polythieno[3,4-b]-thiophene-co-benzodithiophene (PTB7) is a promising electron-donor polymer for organic solar cells.[25] Low band-gap (around 1.6 eV) is obtained via the stabilization of a quinoidal structure from thieno-[3,4-b]thiophene, resulting in a strong absorption around 700 nm, which is the region of the maximum photon flux of the solar spectrum. In addition, it features a rigid backbone that ensures a relatively high hole mobility and environmental stability,[26] whereas the presence of side chains allows a good dispersibility in organic solvents and miscibility with organic acceptors, as fullerene derivatives. As a result of these advantageous properties, a photon-to-current efficiency as high as 7.4% has been reported in OPVs incorporating PTB7.[25, 27, 28] Furthermore, the employment of a solvent mix has been demonstrated to improve the interpenetration of polymers in the blend.[29, 30] In the case of PTB7, the mix of 1,2-Dichlorobenzene ODCB/1,8-diiodoctane (DIO) (97%-3% by volume) has proven to increase the fill factor in the I-V curve.[27] The photophysics in PTB7:PCBM blends has been widely studied [24–30],[31-37] and it is characterised by an ultrafast cation state formation via intra- and intermolecular charge separation, and the formation of polaronic states that is facilitated by the charge transfer character of the repeated units. Differently from P3HT, optical charge separation occurs within very small (2-20 nm) domain sizes.[31]

Here, we extend and nicely complement our previous studies on P3HT:SWNTs blends[24] with a novel investigation on SWNTs in close contact with PTB7 and PCBM. We observe evidence of a hole transfer process between SWNTs to polymer chains, thus demonstrating that SWNTs can participate actively in the exciton dissociation process of such photovoltaic blends.

## 2    Experimental section

### 2.1    Materials

The employed glass substrates were pre-cleaned by means of sonication in acetone and isopropanol and dried out under nitrogen flow. After that a 10 minutes $O_2$ plasma was used to clean the substrates before the deposition of the organic layers. The substrates were then transferred into the glovebox. For the active layer preparation, PTB7 and PC60BM were dissolved in 1,2-Dichlorobenzene (ODCB) (ratio 1:0.8) with concentrations of 20 mg/mL and 16 mg/mL, respectively. The (6,5) SWNTs for the bilayers structures were dispersed as purchased in N, N-Dimethylformamide (DMF), sonicated for 1 hour and centrifuged at 1200 rpm for 30 minutes. The supernatant was then collected. For the bulk and for the bare carbon nanotube samples, the SWNTs were dispersed in ODCB and tip sonicated for 1.5 hours. The active layer was spin-coated at 1200 rpm, yielding a thickness of about 80 nm as measured by profilometry. For the structure PTB7: PCBM/SWNT, a relatively brief oxygen plasma treatment (5 seconds of exposure) was performed to the active layer before deposition of the SWNTs, with the aim to improve DMF wettability on the active layer.

### 2.2    Optical measurements

Steady state absorption spectra have been acquired with a Perkin Elmer spectrophotometer Lambda 1050 WB. Ultrafast differential transmission measurements have been performed with a Ti:Sapphire amplified laser system (repetition rate of 1 kHz, pulse duration of about 100 fs, central wavelength at 780 nm). Excitation pulses have been obtained with second harmonic generation or by optical parametric amplification.[38] Chirp-free transient differential transmission (ΔT/T) has been collected by using an optical multichannel analyser (OMA) with a dechirping algorithm.

## 3  Results and discussion

To elucidate the role of SWNTs in the exciton dynamics of PTB7:PCBM blends (see figure 1b for the chemical structure of the polymer), we employed ultrafast pump-probe spectroscopy to study the photophysics of three sample configurations, characterised by a different interfacial arrangement between the PTB7:PCBM blends and the nanotubes, namely: **1.** (6,5) SWCNT films are deposited below or **2.** on top of the bulk heterojunction layer and **3.** SWNT:PTB7:PCBM heterojunction blend (figure 1c). The steady-state UV-VIS optical absorptions of the three configurations are shown in Figure 1d. We can clearly observe the two vibronic peaks of PTB7 at 625 and 680 nm, while the presence of (6,5) SWNTs in testified by the peak in the infrared region. The relative increase of the polymer vibronic peak at 625 nm suggests that the presence of (6,5) SWNTs leads to an enhancement of the degree of order of the polymer, as it has been already observed for blends of P3HT:SWNTs in previous investigations.[39-41]

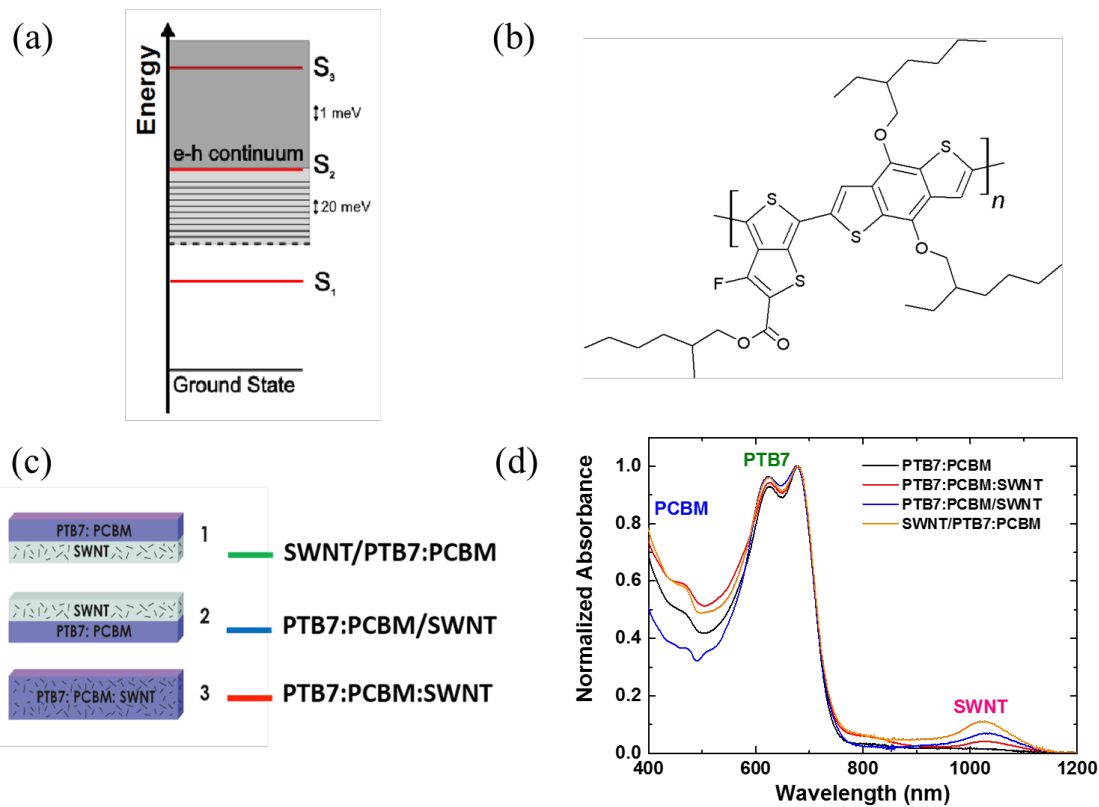

**Figure 1**. (a) Energy diagram landscape of semiconducting carbon nanotubes. (b) Chemical structure of PTB7. (c) The three different sample configurations studied for PTB7:PCBM blend with addition of (6,5) SWNTs. (d) Evolution of the steady state absorption spectra of the PTB7:PCBM samples with three different configurations upon addition of the (6,5) SWNTs.

We highlighted the contribution of SWNTs to the overall deactivation dynamics by pumping selectively either the polymer:fullerene system at 400 nm or the (6,5) SWNTs first exciton $S_1$ at 870 nm, and probing both in the UV-VIS (450-750 nm) and IR (900-1200 nm) spectral regions. The normalised differential transmissions (excitation at 400 nm) in the visible region for the three configurations are reported in Figure 2a for time-delay of 1 ps. In this case, in which we mostly excite and probe the polymer:fullerene system, we can observe that the signal is dominated by the ground-state bleaching (PB) of PTB7. The transient photodynamics of the polymer probed at 680 nm (figure 2b) shows a faster dynamics for the bilayer structures containing SWNTs with respect to the bare PTB7:PCBM heterojunction blend, an effect that can be attributed to an enhanced excitonic relaxation in the presence of SWNTs. In these regards, it has been reported that SWNTs can act as dissociation centres and minimize the radiative recombination at a polymer/SWNT interface.[42] Instead, the slower dynamic in the ternary SWNT:PTB7:PCBM blend can be ascribed to the formation of SWNTs aggregate disrupt PTB7:PCBM interpenetrated network and, thus, reduce the occurrence of charge dissociation events.

In Figure 2c we show the differential transmission pumping at 870 nm in the region of the first exciton, $S_1$, of the (6,5) SWNTs. As it has been already observed in P3HT:PCBM blends in presence of SWNTs,[24] the PB peaks of PTB7:PCBM/SWNT and for the pure SWNTs largely overlap, while SWNT/PTB7:PCBM and the PTB7:PCBM:SWNT show a slight blue-shift of the peak. This effect can be explained in terms of different SWNTs PTB7:PCBM interfacial architectures. In particular, when we cast the nanotubes on top of the polymer:fullerene blend (structure 2) there is less intermixing than in the case of the bulk heterojunction blend configuration (structure 3). Such morphological scenario is further corroborated by the decay dynamics (figure 2d), which displays a similar decay of the SWNT signal film with the PTB7:PCBM/SWNT one (structure 2), whereas a slower decay is noticed for the PTB7:PCBM:SWNT and for the SWNT/P3HT:PCBM sample (structure 3 and 1, respectively), indicating a more effective intermixing of SWNTs with the polymer:fullerene blend for those latter configurations.

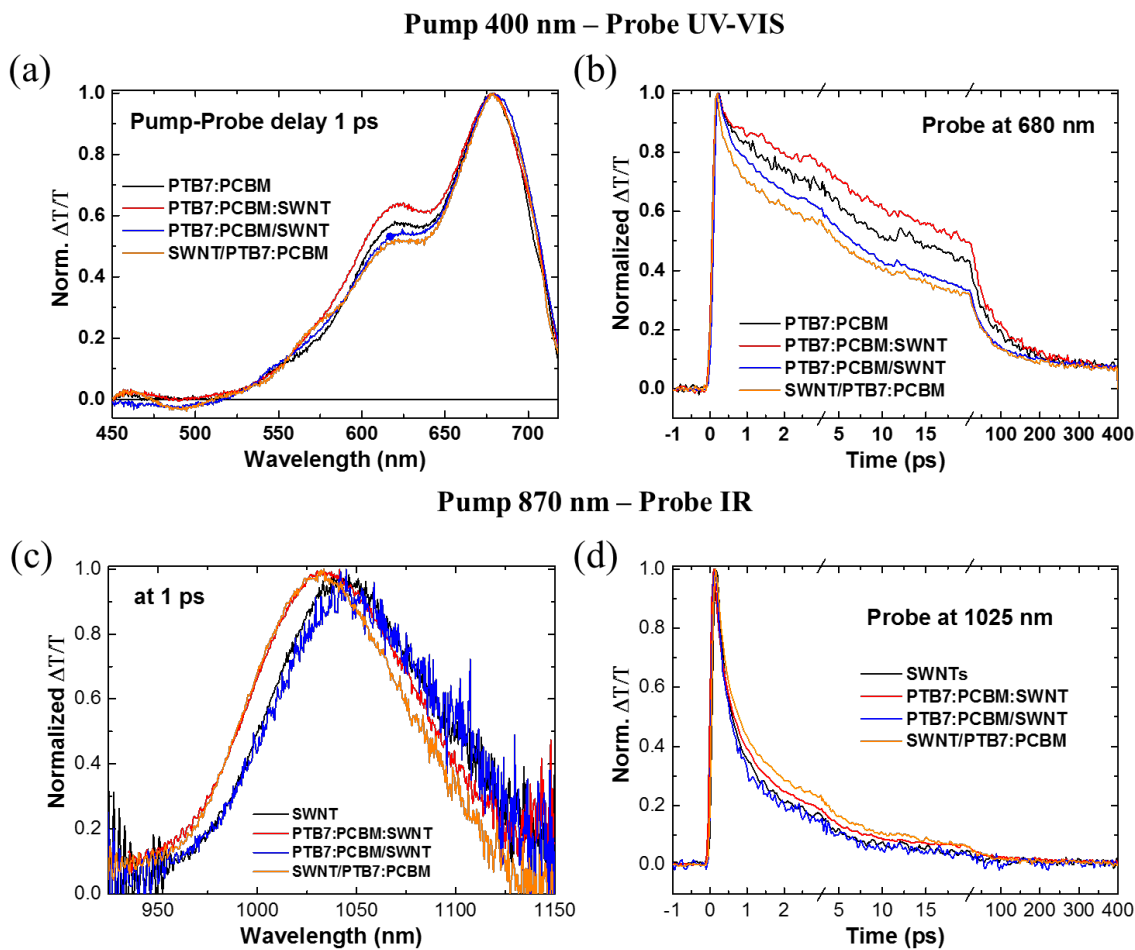

**Figure 2.** (a) Normalised differential transmission spectrum in the UV-VIS region at a pump excitation of 400 nm (pump-probe delay 1 ps). (b) Normalized kinetics of the four samples at a probe wavelength centred at 680 nm, which coincides with the main absorption of the polymer (c) Normalized transient ΔT/T spectra in the IR for a pump excitation at 870 nm (pump-probe delay 1 ps). (d) Normalized dynamics probed at 1025 nm upon selective excitation of the $S_1$ transition.

When exciting the $S_1$ of SWNTs and probing in the visible range (Figure 3a), we observe two PB signals corresponding to the $S_{22}$ transition of the (6,5) SWNT at 580 nm and, in addition, a peak at 680 nm due to the same transition of the (7,5) chirality that is also present in the SWNTs mixture. Interestingly, we note a relatively narrower PB peak from the nanotubes in the bilayers structures, and a slightly red shifted signal for the (6,5) SWNTs, for the heterojunction blend when compared with the pure SWNTs sample. Such differences detected in the transient absorption spectra are also evident from the time-decay plots recorded at probe wavelengths of 590 nm, 680 nm and 690 nm (Figure 3b, c and d). In particular, if we probe at a wavelength corresponding to the $S_2$ transition for the (6,5) SWNTs (580 nm), all the time-decays approach to each other, for the exception of a slight fast decay in the first picosecond observed for the PTB7:PCBM/SWNT consistent with a better dispersion and a higher loading of SWNTs when compared to the bare sample. Similarly to what observed for the (6,5) SWNTs, the (7,5) SWNTs exciton also follows the same decay for all the samples (figure 3c). However, at 690 nm we note a faster deactivation dynamics for the bulk heterojunction structure PTB7:PCBM:SWNT than the other three configurations and pure SWNTs samples. This suggests a possible ultrafast (within 1 ps) charge transfer between PTB7 and SWNTs.

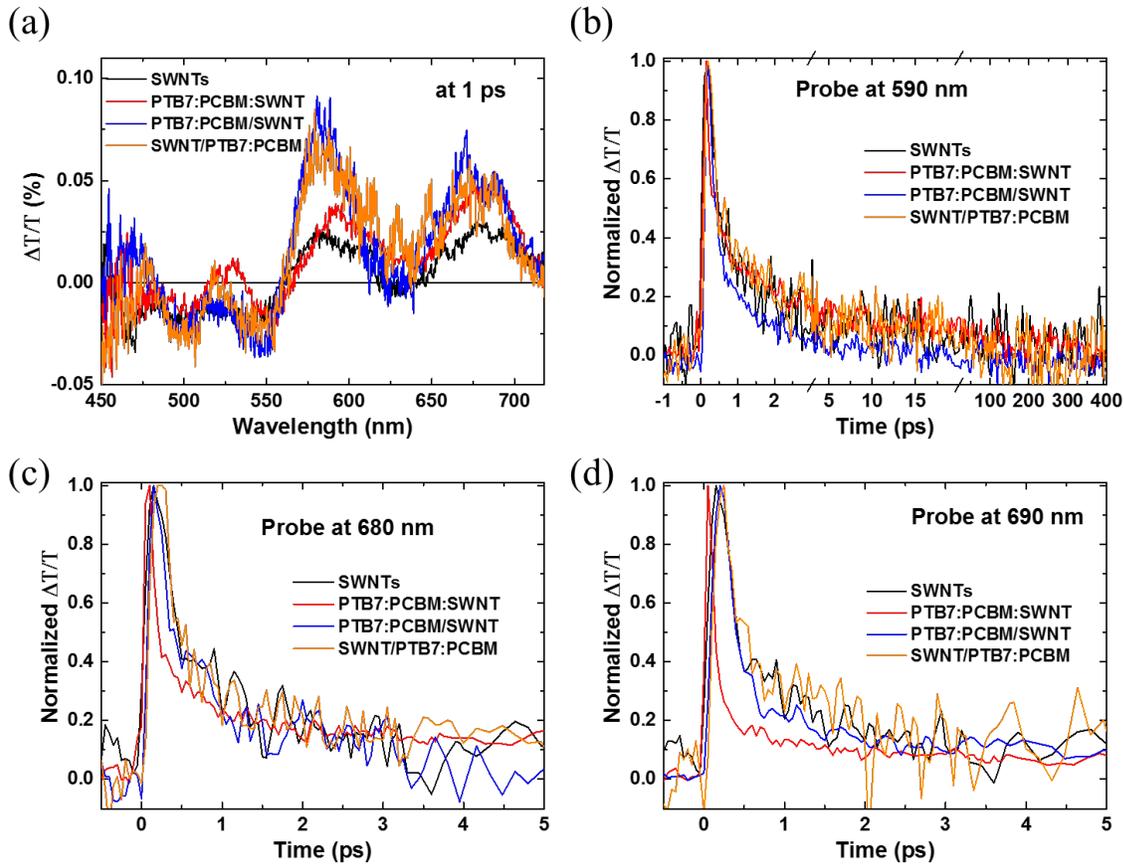

**Figure 3. (a)** Differential transmission spectra for the three different structures studied and bare SWNTs excited at 870 nm and probed in the visible region (delay of 1 ps). Differential transmission dynamics for all the samples probed at (b) 580 nm, (c) 680 nm and (d) 690 nm.

Therefore, to acquire more details on the possible charge transfer mechanism between SWNTs and PTB7, we analysed the building up of the transient spectra within the first picosecond (figure 4a-d, excitation 870 nm). Starting from the pure SWNTs sample consisting of a mixture of (6,5) and (7,5) chiralities (fig. 4a), we see that PB signal at 675 nm due to the (7,5) SWNTs is more intense than the (6,5) at the early pump-probe time-delays, whereas their intensities become comparable after 400 fs. Interestingly, for the PTB7:PCBM:SWNT bulk heterojunction sample (fig. 4b), we see clear evidence of a spectral contribution at 690 nm that can be linked to the polymer PB, which decays almost within our time-resolution (150-250 fs). The observation of the polymer PB signal while resonantly pumping the carbon nanotubes has been also observed for SWNTS:P3HT:PCBM blends,[24] and has been correlated to a hole transfer from the carbon nanotubes to the polymer. It is worth saying that for PBT7 such phenomenon is more difficult to detect and elusive because of the strong overlap of the PB signal of the polymer with the one from the (7,5) SWNTs. If we pass to the bilayer structures (figure 4 c,d), we cannot see evidence of hole transfer from the nanotubes to the polymer for the PTB7:PCBM/SWNT sample, while the higher intermixing achieved for the SWNT/PTB7:PCBM then the abovementioned bilayer structures allow a certain degree of nanotube/polymer:fullerene interaction (fig. 4d), as indicated by the early PTB7 bleaching appearing for time-delays lower than 200 fs. We also studied the ultrafast SWNTS→PBT7 hole-transfer in blends with higher nanotubes loading, to highlight better such an effect (figure S2).

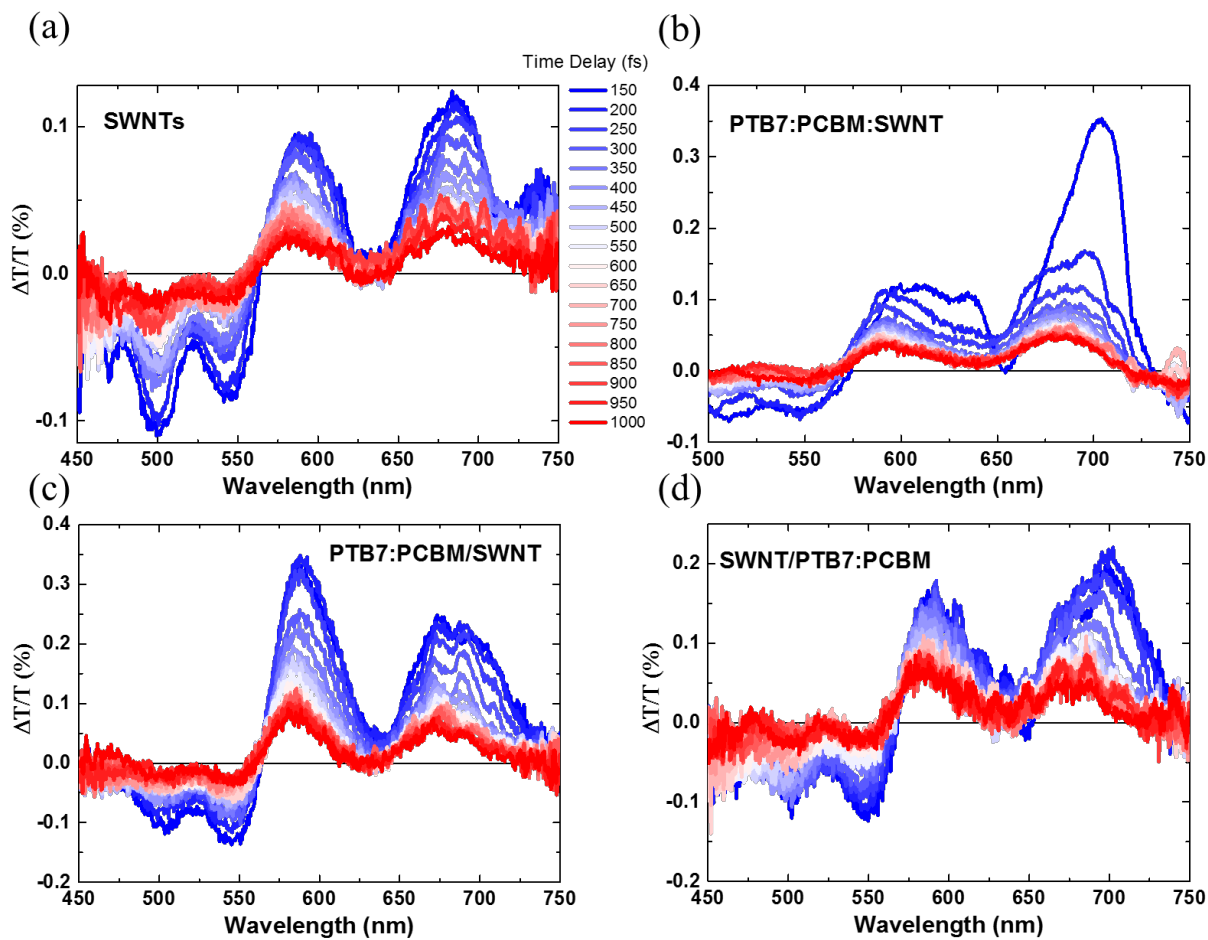

**Figure 4.** Transient differential transmission spectra pumping at 870 nm and probing in the visible region, for pump-probe time delays minor than 1 ps for (a) SWNTs, (b) PTB7:PCBM:SWNT, (c) PTB7:PCBM/SWNT and for (d) SWNT/PTB7:PCBM, respectively.

## 4 Conclusions

In conclusion, we have presented a photophysical study of the behaviour of semiconducting SWNTs placed in close contact with a polymer PTB7 and the fullerene derivative PCBM. We investigated different configurations of this film structures, with the nanotubes as a separate layer above or below PTB7:PCBM, or with the SWNTs in a ternary bulk heterojunction blends. We found evidence of ultrafast SWNTs→PBT7 hole transfer, an effect that is strongly dependent on sample configuration and nanotubes proximity to the polymer-rich regions. These measurements demonstrate an active role of the nanotube in the excitonic photodynamic of such photovoltaic blends and, hence, can be useful for the optimization of photovoltaic nanocomposites integrating SWNTs as active material.

*Conflict of interest*

There are no conflicts of interest to declare.

*Acknowledgments*

G.M.P is supported by the H2020 ETN SYNCHRONICS under a grant agreement 643238.

*Supplementary information for*

# Semiconducting Carbon Nanotubes in photovoltaic blends: the case of PTB7:PC$_{60}$BM:(6,5) SWNT

**Diana Gisell Figueroa del Valle**[a§], **Giuseppe M. Paternò**[b§], **Francesco Scotognella**[b,c]

[a] Instituto Superior de Tecnologias y Ciencias Aplicadas, InSTEC, Ave. Salvador Allende, esq. Luaces, La Habana, Cuba
[b] Center for Nano Science and Technology@PoliMi, Istituto Italiano di Tecnologia, Via Pascoli 70/3, 20133 Milano, Italy
[c] Dipartimento di Fisica, Politecnico di Milano, Piazza L. da Vinci 32, 20133 Milano, Italy

To better observe the bleaching from the polymer while selective pumping the SWNTs, we studied a PTB7:SWNT sample with a higher loading of nanotubes. For this sample a 2:0.5 mg/l polymer:SWNT ratio was employed. The polymer and the SWNT were dispersed in ODCB and stirred overnight at 60 °C. The solution was then ultrasonicated for 1 hour and ultra-centrifugated at 12000 rpm for 15 min, the supernatant was collected and drop cast in a glass substrate. The transient absorption spectra pumped at 870 nm while probed in the visible region is presented in Figure S1.

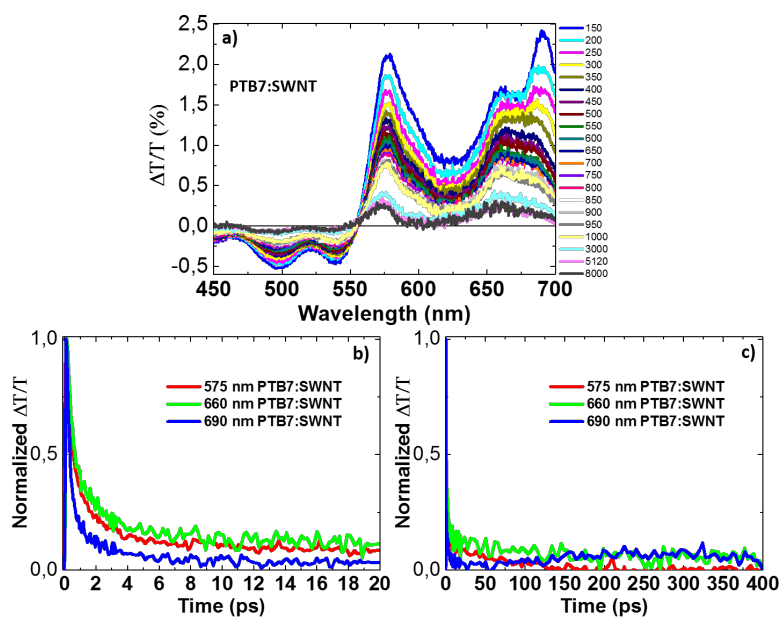

**Figure S1.** a) Transient differential transmission spectra and b-c) dynamics decays for a PTB7:SWNT drop casted sample pumped at 870 nm and probed in the visible region.